\documentclass[review]{elsarticle}

\usepackage{lineno}
\usepackage[colorlinks=true,linkcolor=black,citecolor=blue,urlcolor=blue,pdfauthor=author]{hyperref}
\modulolinenumbers[5]
\usepackage{amsmath,amsfonts,amssymb,amsthm,bm}
\usepackage{float}
\usepackage{graphics}
\usepackage{graphicx}
\usepackage{caption}
\graphicspath{{figures/}}
\usepackage{subfigure}
\usepackage{fullpage}
\usepackage{multirow}
\usepackage{booktabs}
\usepackage{textcomp}
\usepackage{gensymb}
\usepackage{nomencl}
\usepackage{comment}
\usepackage{setspace}
\usepackage{geometry}
\usepackage[section]{placeins}
\usepackage{soul}
\journal{Elsevier}

\begin{document}
\captionsetup[figure]{labelfont={bf},name={Fig.},labelsep=period}
\begin{frontmatter}

\title{Lamb’s problem for a half‐space coupled to a generic distribution of oscillators at the surface}

\author[mymainaddress] {Xingbo Pu \fnref{fn1}}
\author[mymainaddress] {Antonio Palermo\fnref{fn1}}

\author[mymainaddress] {Alessandro Marzani\corref{mycorrespondingauthor}}
\cortext[mycorrespondingauthor]{Corresponding author}
\fntext[fn1]{Equal contributors.}

\ead{alessandro.marzani@unibo.it}

\address[mymainaddress]{Department of Civil, Chemical, Environmental and Materials Engineering, University of Bologna, 40136 Bologna, Italy}

\begin{abstract}

We propose an analytical framework to model the effect of single and multiple mechanical surface oscillators on the dynamics of vertically polarized elastic waves propagating in a semi-infinite medium. The formulation extends the canonical Lamb's problem, originally developed to obtain the wavefield induced by a harmonic line source in an elastic half-space, to the scenario where a finite cluster of vertical oscillators is attached to the medium surface. In short, our approach utilizes the solution of the classical Lamb's problem as Green's function to formulate the multiple scattered fields generated by the resonators. For an arbitrary number of resonators, arranged atop the elastic half-space in an arbitrary configuration, the displacement fields are obtained in closed-form and validated with numerics developed in a two-dimensional finite element environment. 

\end{abstract}


\end{frontmatter}

\linenumbers

\section{Introduction}

Modeling the propagation of mechanical surface waves in an elastic half-space is a long-lasting topic in physics and engineering.
A cornerstone of this research topic is the seminal work by Lamb \cite{lamb1904propagation} which describes the fundamental solution for a harmonic load applied on the surface of an elastic medium, a scenario currently known as the Lamb's problem. Since then, numerous researchers have enriched the complexity of this problem accounting for the presence of inclusions, obstacles, profile and material discontinuities along and within the elastic medium \cite{luco1994dynamic, tanaka1998surface, malfanti2011propagation, brule2014experiments, li2018non, wang2019scattering, zhao2020non}.\par

A canonical problem of particular interest concerns the propagation of elastic waves in a semi-infinite substrate supporting a cluster of resonant elements. This configuration can indeed illustrate problems of technological relevance across different length-scales, as seismic waves interacting with the built environment \cite{jennings1973dynamics, boutin2006wave, ghergu2009structure} or surface waves propagating in micro-mechanical resonant systems \cite{khelif2010locally, boechler2013interaction}. Additionally, periodic clusters of surface resonators have been recently explored to realize novel devices for surface wave manipulation, the so-called elastic metasurfaces. Among these periodic configurations, arrays of beams or pillars \cite{colquitt2017seismic, wootton2019asymptotic, chaplain2020tailored}, and mass-spring resonators \cite{garova1999interaction, boechler2013interaction, pu2020seismic} have shown the capabilities to shape both the direction of propagation and the frequency content of elastic waves. Pivotal in all these coupled substrate-resonators engineering problems is the knowledge of both  dispersion relation and wavefield.\par

Several analytical formulations are currently available to derive the dispersive properties \cite{colquitt2017seismic} and transmission coefficients of metasurfaces \cite{boutin2006wave, marigo2020effective}. In most cases, these approaches describe the collective behavior of an infinite array of oscillators with the aid of an effective medium approach \cite{boechler2013interaction, maznev2015waveguiding}, or via asymptotic and homogenization techniques \cite{boutin2006wave, marigo2020effective, schwan2013unconventional}.\par 
The calculation of the elastic wavefield of a finite-size, arbitrarily distributed cluster of resonators is instead obtained via numerical techniques (like standard \cite{PALERMO2020104181} or spectral FEM \cite{colombi2016seismic}), since no closed-form formulation is currently available to this purpose. Nonetheless, only the knowledge of the wavefield can shed light on the destructive or constructive wave interference generated by the resonators array which is in turn responsible for peculiar wave phenomena like surface-to-bulk wave conversion \cite{colquitt2017seismic, pu2020seismic}, rainbow trapping and wave localization \cite{colombi2016seismic, chaplain2020topological}. Despite the possibility to obtain actual results for specific configurations by means of numerical schemes, analytical treatment of this elastodynamic problem can allow (i) to better comprise the nature of these phenomena, (ii)  to guide the optimal design of waves control devices, and (iii) to derive general conclusions on the interaction problem between closed resonators mechanically coupled by an elastic substrate. \par

Hence, in this work we develop an exact formulation which extends the classical Lamb's problem to the case of an elastic half-space coupled to an arbitrary cluster of vertical surface resonators. To this purpose, we calculate the incident wavefield generated by a harmonic source following the approach by Lamb. The Lamb's solution is also used as Green's function to describe the scattered field generated by each resonator when excited by a harmonic motion at its base. The overall substrate wavefield is then obtained as solution of the coupled problem due to the interference of the incident field and the multiple scattered fields of the oscillators. Our formulation can tackle a generic number of different resonators located at arbitrary distances from the source, as illustrated in the various examples discussed in the work and validated against numerical results (FEM). \par

The article is organized as follows. In Section \ref{Analytical framework}, we present our analytical formulation. We begin by describing the solution of the Lamb's problem for a harmonic source applied on the free surface of a semi-infinite elastic substrate (Section \ref{Surface harmonic load on a half-space}). Then, we recall the response for an oscillator subjected to a harmonic motion at its base (Section \ref{Dynamics of surface resonators}) and formulate the interaction problem between resonators and the half-space (Section \ref{Surface resonators coupled to the half-space}). In Section \ref{Case studies}, we calculate the response of an elastic substrate with a single, a pair and a cluster of surface resonators and validate our predictions against numerics. Finally, in Section \ref{Conclusions} we summarize the main findings of our work.

\section{Analytical framework} \label{Analytical framework}

\begin{figure}[hbt!]
    \centering
    \includegraphics[width=5 in]{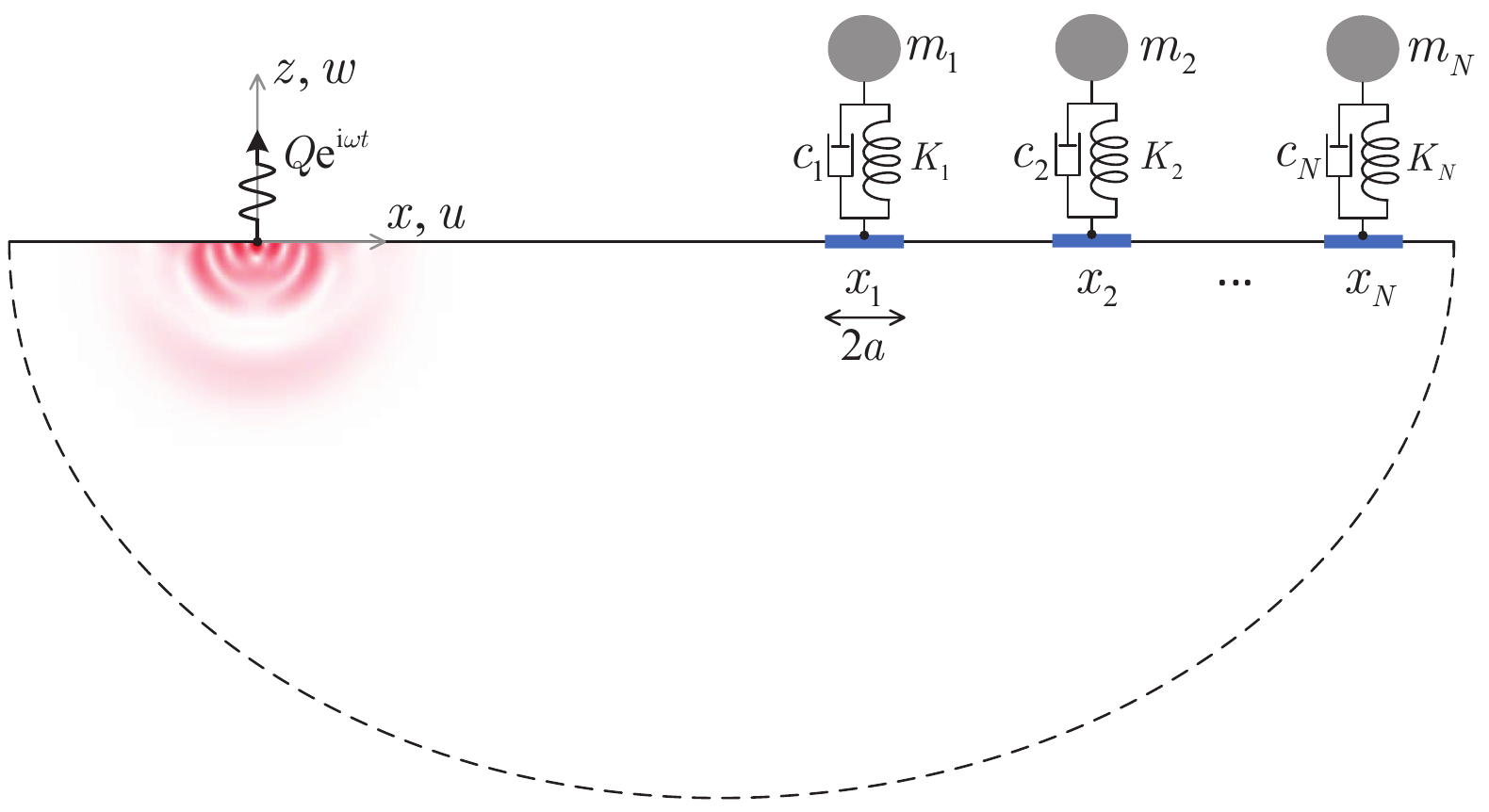}
    \caption{Schematic of Rayleigh wave interacting with resonators on an elastic half-space.}
    \label{fig:fig1}
\end{figure}

We develop an analytical framework to calculate the response of an isotropic, linear elastic half-space coupled with $N$ oscillators and excited by a harmonic line source  (see Fig. \ref{fig:fig1}).

Our investigation begins  recalling (i) the solution of the Lamb's problem for a harmonic line load applied at the free surface of an elastic, isotropic half-space \cite{lamb1904propagation} and (ii) the response of a vertical oscillator to an imposed harmonic base motion. The response of the coupled system is obtained by formulating the interaction problem between the source-generated wavefield, the solution of the Lamb's problem, and the summation of the scattered wavefields generated by the motion of the surface resonators. 

\subsection{Surface harmonic load on a half-space} \label{Surface harmonic load on a half-space}

Let us consider a time-harmonic force per unit length $Q\mathrm{e}^{\mathrm{i}\omega t}$  applied normal to the free surface of an isotropic elastic medium. We resort to a two-dimensional (2D) plane-strain formulation in the $x-z$ plane, where the $x$-axis is directed along the wave propagation and the $z$-axis is  perpendicular to the free surface (see  Fig. \ref{fig:fig1}). In absence of resonators, free stress boundary conditions ($\sigma_{zz}=\tau_{zx}=0$) characterize the elastic half-space along its whole surface except for the source location where:
\begin{equation} \label{equ:B.C for Lamb problem}    
    \sigma_{zz} (x, z, t) = Q\delta(x)  \mathrm{e}^{\mathrm{i}\omega t}, \quad  
    \tau_{zx} (x, z, t) = 0, \quad \text{for} \quad x=0,\; z=0
\end{equation}

Given the plane-strain conditions, the displacement vector lies in the plane $x-z$ with non-null components denoted as $\bm{u}(x,z,t)=[u, w]$. It is here convenient to use the Helmholtz decomposition $\bm{u}=\nabla \Phi+\nabla \times \bm{\Psi}$ and to express the displacement components as:

\begin{equation} \label{equ:displacement expression}    
    u=\frac{\partial \Phi}{\partial x}-\frac{\partial \Psi_{y}}{\partial z}, \quad  w=\frac{\partial \Phi}{\partial z}+\frac{\partial \Psi_{y}}{\partial x}
\end{equation}
where $\Phi(x,z,t)$ is a scalar potential and $\Psi_{y}(x,z,t)$ is a component of the vector potential $\bm{\Psi}$. \par

Restricting our interest to the steady-state condition, the potentials assume the form:
\begin{equation}
    \Phi(x,z,t)=\Phi(x,z)\mathrm{e}^{\mathrm{i}\omega t}, \quad \Psi_y(x,z,t)=\Psi_y(x,z)\mathrm{e}^{\mathrm{i}\omega t}
\end{equation}
and satisfy the wave equations  \cite{achenbach1973wave}:
\begin{equation} \label{equ:wave equation}
    \nabla^{2} \Phi + k_p^{2} \Phi=0,  \quad  \nabla^{2} \Psi_y + k_s^{2} \Psi_y=0
\end{equation}
\noindent
in which $k_{p}$ and $k_{s}$  denote, respectively, the wave number of pressure and shear waves in the substrate, namely:

\begin{equation} 
    k_p=\frac{\omega}{c_p}, \quad k_s=\frac{\omega}{c_s}
\end{equation}
\noindent
where:
\begin{equation} 
     c_p=\sqrt{\frac{\lambda+2 \mu}{\rho}}, \quad c_s=\sqrt{\frac{\mu}{\rho}}
\end{equation}
are the pressure and shear wave velocities, respectively, $\lambda$ and $\mu$  the Lam\'e constants and $\rho$  the mass density of the substrate. 

According to the Hooke's law and employing the Helmholtz decomposition in Eq. \eqref{equ:displacement expression}, the in-plane components of the stress tensor $\bm{\sigma}$ can be expressed as function of the potentials:

\begin{equation} \label{equ:stress tensor}
    \sigma_{zz}=-\mu \left[k_s^{2} \Phi + 2\left(\frac{\partial^2 \Phi}{\partial x^2} -\frac{\partial^2 \Psi_y}{\partial x \partial z}\right)\right], \quad
    \tau_{zx}=-\mu \left[k_s^{2} \Psi_y - 2\left(\frac{\partial^2 \Phi}{\partial x \partial z} -\frac{\partial^2 \Psi_y}{\partial z^2}\right)\right]
\end{equation}

At this stage, we seek for the solutions of the wave Eqs.  \eqref{equ:wave equation} by means of the Fourier transform along the $x$-direction:
\begin{equation} \label{equ:Fourier transform to wave equation}
    \mathcal{F} \{\nabla^{2} \Phi + k_p^{2} \Phi\}=\frac{\partial^2 \bar{\Phi}}{\partial z^2}-(k^2-k_p^{2})\bar{\Phi}=0,  \quad  
    \mathcal{F} \{\nabla^{2} \Psi_y + k_s^{2} \Psi_y\}=\frac{\partial^2 \bar{\Psi}_y}{\partial z^2}-(k^2-k_s^{2})\bar{\Psi}_y=0
\end{equation}
\noindent which admit general solutions of the form:
\begin{equation} \label{equ:Fourier solution of wave equation}
    \bar{\Phi} (k, z)=A_{1} \mathrm{e}^{-pz}+B_{1} \mathrm{e}^{pz}, \quad
    \bar{\Psi}_{y} (k, z)=A_{2} \mathrm{e}^{-qz}+B_{2} \mathrm{e}^{qz}
\end{equation}
\noindent
with:
\begin{equation}
    p=\sqrt{k^2-k_p^2}, \quad q=\sqrt{k^2-k_s^2}
\end{equation}
\noindent
and where the coefficients $A_1$ and $A_2$ in Eq. (\ref{equ:Fourier solution of wave equation}) must be equal to zero to avoid unbounded responses at increasing depth $z$. The remaining coefficients $B_1$ and $B_2$ are determined by imposing the boundary conditions. \par

Fourier transforming the stress components in Eq. (\ref{equ:stress tensor}), and making use of the boundary conditions in Eq. (\ref{equ:B.C for Lamb problem}), lead to the following expressions:

\begin{subequations}

\begin{equation} \label{equ:B.C lamb fourier a}
     (2k^2-k_{s}^2) B_1 + 2\mathrm{i}kq B_2 = Q/\mu 
\end{equation}
\begin{equation} \label{equ:B.C lamb fourier b}
     -2\mathrm{i}kp B_1 + (2k^2-k_{s}^2) B_2 = 0 
\end{equation}
\end{subequations}
\noindent
Solutions of Eqs. \eqref{equ:B.C lamb fourier a} and \eqref{equ:B.C lamb fourier b} provide the coefficients:

\begin{equation} 
     B_1 = \frac{Q}{\mu} \frac{(2k^2-k_{s}^2)}{R(k)}, \quad
     B_2 = \frac{Q}{\mu} \frac{2\mathrm{i}kp}{R(k)}
\end{equation}
\noindent
where $R(k)$ denotes the so-called Rayleigh function:
\begin{equation} 
     R(k) \equiv (2k^2-k_s^2)^2-4k^2 p q 
\end{equation}
\noindent
The inverse Fourier transform of Eq. (\ref{equ:Fourier solution of wave equation}) provides the expression of the potentials in the plane $x-z$:

\begin{subequations}
\begin{equation} \label{equ:solution of phi}
     \Phi(x, z)=\mathcal{F}^{-1} \{\bar{\Phi}\}=\frac{Q}{2\pi\mu} \int_{-\infty}^{\infty}\frac{2k^2-k_s^{2}}{R(k)} \mathrm{e}^{pz+\mathrm{i}kx}\,\mathrm{d}k
\end{equation}
\begin{equation} \label{equ:solution of psi}
     \Psi_{y}(x, z)=\mathcal{F}^{-1} \{\bar{\Psi}_{y}\}=\frac{Q}{2\pi\mu} \int_{-\infty}^{\infty}\frac{2\mathrm{i}kp}{R(k)} \mathrm{e}^{qz+\mathrm{i}kx}\,\mathrm{d}k 
\end{equation}
\end{subequations}
\noindent
At last, by substituting Eqs. (\ref{equ:solution of phi}) and (\ref{equ:solution of psi}) into Eq. (\ref{equ:displacement expression}), the displacement components of the wavefield induced by the time-harmonic line load are obtained as:

\begin{subequations}
\begin{equation} \label{equ:general solution of u}
     u^{(f)}(x,z)=\frac{\mathrm{i}Q}{2\pi\mu} \int_{-\infty}^{\infty}\frac{k(2k^2-k_s^{2}) \mathrm{e}^{pz}-2kpq \mathrm{e}^{qz}}{R(k)} \mathrm{e}^{\mathrm{i}kx}\,\mathrm{d}k  
\end{equation}
\begin{equation} \label{equ:general solution of w}
     w^{(f)}(x,z)=\frac{Q}{2\pi\mu} \int_{-\infty}^{\infty}\frac{p(2k^2-k_s^{2}) \mathrm{e}^{pz}-2k^2p \mathrm{e}^{qz}}{R(k)} \mathrm{e}^{\mathrm{i}kx}\,\mathrm{d}k
\end{equation}
\end{subequations}
\noindent
where the superscript ($f$) is used to label these displacement components of the free field (no resonators). From Eqs. \eqref{equ:general solution of u} and \eqref{equ:general solution of w} the free field displacement components at $z=0$ can be expressed as \cite{lamb1904propagation}:

\begin{subequations}
\begin{equation} \label{equ:solution of u at z=0}
     u^{(f)}(x,0)=\frac{\mathrm{i}Q}{2\pi\mu} \int_{-\infty}^{\infty}\frac{k(2k^2-k_s^{2}-2pq)}{R(k)} \mathrm{e}^{\mathrm{i}kx}\,\mathrm{d}k  
\end{equation}
\begin{equation} \label{equ:solution of w at z=0}
     w^{(f)}(x,0)=-\frac{Q}{2\pi\mu}  \int_{-\infty}^{\infty}\frac{k_s^{2}p}{R(k)}  \mathrm{e}^{\mathrm{i}kx}\,\mathrm{d}k 
\end{equation}
\end{subequations}

\subsection{Dynamics of surface resonators} \label{Dynamics of surface resonators}

We now consider the steady-state dynamics of mass-spring-dashpot resonators located atop an elastic half-space under harmonic motion. The set $\mathcal{O}=\{x_1, x_2, \cdots, x_N \mid N \in \mathbb{Z}^+\} \subset \mathbb{R}$ is introduced to collect the $x$-coordinate of the resonators. For each resonator we identify a footprint area $S$ with a length $2a$ along the $x$-direction.

The dynamic equilibrium equation of each resonator reads:

\begin{equation} \label{equ:motion of resonator}
    m_n \ddot{W_n} + c_n(\dot{W_n}-\dot{\tilde{w}}(x_n,0)) + K_n(W_n-\tilde{w}(x_n,0)) = 0 \quad \text{for} \quad x_n \in \mathcal{O}
\end{equation}
\noindent
where $m_n$, $c_n$ and $K_n$ are the $n$-th resonator mass, viscous damping coefficient and spring stiffness, $W_n$ is the absolute vertical displacement of the $n$-th mass, and $\tilde{w}(x_n,0)$ is the average vertical displacement of the resonator footprint: 

\begin{equation} \label{equ:equation of resonator motion}
    \tilde{w}(x_n,0) =\frac{1}{2a}\int_{x_n-a}^{x_n+a} w(x,0)\,\mathrm{d}x \quad \text{for} \quad x_n \in \mathcal{O}
\end{equation}

The use of an average base displacement is motivated by both physical and mathematical arguments. From a physical point, the average displacement represents the mean motion at the finite-size base of the oscillator. Mathematically, it allows to eliminate the divergence of the Green's function at the origin.

According to Eq. \eqref{equ:equation of resonator motion}, the absolute vertical displacement of the generic $n$-th resonator excited by a harmonic base motion $\tilde{w}(x_n,0)$ of circular frequency $\omega$ reads:

\begin{equation} \label{equ:relation of displacement}
    W_n=\frac{m_n\omega_{rn}^2+\mathrm{i}\omega c_n}{m_n(\omega_{rn}^2-\omega^2)+\mathrm{i}\omega c_n} \tilde{w}(x_n,0)\equiv T_{Rn} \tilde{w}(x_n,0) \quad \text{for} \quad x_n \in \mathcal{O}
\end{equation}
\noindent
where $T_{Rn}$ denotes the so-called transmissibility \cite{chopra2012dynamics} of a damped resonator, and where $\omega_{rn} = \sqrt{K_n / m_n}$ is the angular resonant frequency of the $n$-th resonator. Accordingly, the normal force applied by the resonator to the substrate can be written as:

\begin{equation} \label{equ:normal total stress}
F_n= m_n \omega^2 W_n =\frac{m_n\omega^2(m_n\omega_{rn}^2+\mathrm{i}\omega c_n)}{m_n(\omega_{rn}^2-\omega^2)+\mathrm{i}\omega c_n} \tilde{w}(x_n,0) \equiv \Omega_n \tilde{w}(x_n,0)   \quad \text{for}\quad x_n \in \mathcal{O}
\end{equation}  
\noindent
and the uniform stress exerted by each resonator over the contact area reads:

\begin{equation} \label{equ:B.C resonator}    
    \sigma_{zz} (x,0) = \frac{F_n}{S} =\frac{\Omega_n \tilde{w}(x_n,0)}{2a}, \quad  
    \tau_{zx} (x, 0) = 0, \quad \text{for} \quad  x\in ( x_n-a,x_n+a),\; x_n \in \mathcal{O}
\end{equation}

These harmonic normal stresses behave as sources of additional wavefields in the half-space and interact with the free field generated by the source. The nature and implication of this interaction is described in the next section.  

\subsection{Surface resonators coupled to the half-space} \label{Surface resonators coupled to the half-space}

As anticipated in the previous section, the resonators excited by a harmonic base motion generate additional wavefields in the half-space. We label the $j$-th resonator-induced wavefield $\bm{u}_j^{(s)}(x,z)$, where the superscript $(s)$ is used to denote scattered field, so that the total displacement field of the coupled problem can be written as:
\begin{equation} \label{equ:definition of total field}
    \bm{u}(x,z)=\bm{u}^{(f)}(x,z)+\sum_{j=1}^{N} \bm{u}_{j}^{(s)}(x,z) \equiv \bm{u}^{(f)}(x,z)+\bm{u}^{(s)}(x,z)
\end{equation}
\noindent
\noindent where $\bm{u}^{(s)}(x,z)$ is the total wavefield. The free wavefield $\bm{u}^{(f)}(x,z)$, except for the point of application of the harmonic force ($x=0$), is characterized by null stress components at the surface:
\begin{equation} 
    \sigma_{zz}^{(f)}(x,0)=\tau_{zx}^{(f)}(x,0)=0  \quad \text{for} \quad x \in \mathbb{R} \setminus \{0\}
\end{equation}
\noindent
hence, the stress at each resonator footprint depends only on the scattered wavefield, namely:
\begin{subequations}
\begin{equation} \label{equ:normal stress at x_j}
    \sigma_{zz}(x,0)=\sigma_{zz}^{(f)}(x,0)+ \sigma_{zz}^{(s)}(x,0)=\sigma_{zz}^{(s)}(x,0)  \quad \text{for} \quad x\in ( x_n-a,x_n+a),\; x_n \in \mathcal{O}
\end{equation}
\begin{equation} \label{equ:shear stress at x_j}
    \tau_{zx}(x,0)=\tau_{zx}^{(f)}(x,0)+ \tau_{zx}^{(s)}(x,0)=0   \quad \text{for} \quad x\in ( x_n-a,x_n+a),\; x_n \in \mathcal{O}
\end{equation}
\end{subequations}
\noindent
In force of Eq. \eqref{equ:B.C resonator} and Eq. \eqref{equ:definition of total field}, the scattered normal stress at the resonator footprint in Eq. \eqref{equ:normal stress at x_j} can be written as:

\begin{equation} \label{equ:normal scattered stress}
    \sigma_{zz}^{(s)}(x,0)=\frac{\Omega_n \tilde{w}(x_n,0)}{2a}= \frac{\Omega_n[\tilde{w}^{(f)}(x_n,0)+\tilde{w}^{(s)}(x_n,0)]}{2a}   \quad \text{for} \quad x\in ( x_n-a,x_n+a),\; x_n \in \mathcal{O}
\end{equation}
where the average free-field vertical displacement $\tilde{w}^{(f)}(x_n,0)$ at the footprint of the $n$-th resonator can be obtained exploiting Eq. \eqref{equ:general solution of w} as:

\begin{equation}
\label{equ:free w at x_n}
   \tilde{w}^{(f)}(x_n,0)=\frac{1}{2a} \frac{-Q}{2\pi\mu} \int_{x_n-a}^{x_n+a} \int_{-\infty}^{\infty}\frac{k_s^{2} p}{R(k)} \mathrm{e}^{\mathrm{i}kx}\,\mathrm{d}x \mathrm{d}k = \frac{-Q}{2\pi\mu} \int_{-\infty}^{\infty}\frac{k_s^{2} p}{R(k)} \frac{\sin(ka)}{ka} \mathrm{e}^{\mathrm{i}kx_n}\,\mathrm{d}k \quad \text{for} \quad x_n \in \mathcal{O}
\end{equation}

The  average scattered field $\tilde{w}^{(s)}(x_n,0)$ can be instead obtained as described in the following.
First Eqs. \eqref{equ:general solution of u} and \eqref{equ:general solution of w} are used as Green's functions to express the displacement components of the scattered field $\bm{u}_j^{(s)}(x,z)$ generated by the normal stress exerted by the $j$-th resonator: 

\begin{subequations}
\begin{equation} \label{equ:scattered u by the j-th resonator}
\begin{split}
    u_j^{(s)}(x,z) &=\mathrm{i}\int_{x_j-a}^{x_j+a} \frac{\sigma_{zz}^{(s)}(\eta,0)}{2\pi\mu} \int_{-\infty}^{\infty} \frac{k(2k^2-k_s^{2}) \mathrm{e}^{pz}-2kpq \mathrm{e}^{qz}}{R(k)} \mathrm{e}^{\mathrm{i}k(x-\eta)}\,\mathrm{d}\eta \mathrm{d}k \\
    & = \frac{\mathrm{i}\Omega_j [\tilde{w}^{(f)}(x_j,0)+\tilde{w}^{(s)}(x_j,0)]}{2\pi\mu} \int_{-\infty}^{\infty}\frac{k(2k^2-k_s^{2}) \mathrm{e}^{pz}-2kpq \mathrm{e}^{qz}}{R(k)} \frac{\sin(ka)}{ka} \mathrm{e}^{\mathrm{i}k(x-x_j)}\,\mathrm{d}k   \quad \text{for} \quad x_j \in \mathcal{O}
\end{split}
\end{equation}
\begin{equation} \label{equ:scattered w by the j-th resonator}
\begin{split}
    w_j^{(s)}(x,z) &=\int_{x_j-a}^{x_j+a} \frac{\sigma_{zz}^{(s)}(\eta,0)}{2\pi\mu} \int_{-\infty}^{\infty} \frac{p(2k^2-k_s^{2}) \mathrm{e}^{pz}-2k^2p \mathrm{e}^{qz}}{R(k)} \mathrm{e}^{\mathrm{i}k(x-\eta)}\,\mathrm{d}\eta \mathrm{d}k \\
    & = \frac{\Omega_j [\tilde{w}^{(f)}(x_j,0)+\tilde{w}^{(s)}(x_j,0)]}{2\pi\mu} \int_{-\infty}^{\infty}\frac{p(2k^2-k_s^{2}) \mathrm{e}^{pz}-2k^2p \mathrm{e}^{qz}}{R(k)} \frac{\sin(ka)}{ka} \mathrm{e}^{\mathrm{i}k(x-x_j)}\,\mathrm{d}k   \quad \text{for} \quad x_j \in \mathcal{O}
\end{split}
\end{equation}
\end{subequations}
\noindent
Note that the calculation of each scattered wavefield $u_j^{(s)}(x,z)$, and $w_j^{(s)}(x,z)$, as per Eqs. (\ref{equ:scattered u by the j-th resonator}) and (\ref{equ:scattered w by the j-th resonator}), requires only the knowledge of the average vertical displacement,  free  $\tilde{w}^{(f)}(x_j,0)$ and scattered $\tilde{w}^{(s)}(x_j,0)$,  at the contact points $x_j \in \mathcal{O}$. \par 
Since the free field component $\tilde{w}^{(f)}(x_j,0)$ is known from Eq. (\ref{equ:free w at x_n}), the  only left unknowns in Eqs. \ref{equ:scattered u by the j-th resonator} and  \ref{equ:scattered w by the j-th resonator} are the vertical scattered displacements  $\tilde{w}^{(s)}(x_j,0)$.
To calculate them, we  expand the average vertical scattered displacement field at the contact points $x_n$ according to Eq. (\ref{equ:definition of total field}):
\begin{equation}
\tilde{w}^{(s)}(x_{n}, 0)=\sum_{j=1}^{N} \tilde{w}_{j}^{(s)}(x_{n},0) \quad \text{for} \quad n=1,2,\cdots,N
\end{equation}
which can be equivalently rewritten as:

\begin{equation} \label{equ:vertical scattered field}
    \begin{cases}
     \tilde{w}^{(s)}(x_{1}, 0) =\tilde{w}_{1}^{(s)}(x_{1}, 0)+\tilde{w}_{2}^{(s)}(x_{1}, 0)+\cdots+\tilde{w}_{N}^{(s)}(x_{1}, 0) \\
     \tilde{w}^{(s)}(x_{2}, 0) =\tilde{w}_{1}^{(s)}(x_{2}, 0)+\tilde{w}_{2}^{(s)}(x_{2}, 0)+\cdots+\tilde{w}_{N}^{(s)}(x_{2}, 0) \\
     \quad \quad \vdots  \\
     \tilde{w}^{(s)}(x_{N}, 0) =\tilde{w}_{1}^{(s)}(x_{N}, 0)+\tilde{w}_{2}^{(s)}(x_{N}, 0)+\cdots+\tilde{w}_{N}^{(s)}(x_{N}, 0) \\
   \end{cases}
\end{equation}
\noindent
Then, we substitute Eq. (\ref{equ:scattered w by the j-th resonator}) into Eq. (\ref{equ:vertical scattered field}) and obtain the following equations:
\begin{small}
\begin{equation} \label{equ:expanding vertical scattered field}
    \begin{cases}
     \tilde{w}^{(s)}(x_{1}, 0) =\beta_{11}[\tilde{w}^{(f)}(x_{1}, 0)+\tilde{w}^{(s)}(x_{1}, 0)]+\beta_{12}[\tilde{w}^{(f)}(x_{2}, 0)+\tilde{w}^{(s)}(x_{2}, 0)]+\cdots+\beta_{1N}[\tilde{w}^{(f)}(x_{N}, 0)+\tilde{w}^{(s)}(x_{N}, 0)] \\
     \tilde{w}^{(s)}(x_{2}, 0) =\beta_{21}[\tilde{w}^{(f)}(x_{1}, 0)+\tilde{w}^{(s)}(x_{1}, 0)]+\beta_{22}[\tilde{w}^{(f)}(x_{2}, 0)+\tilde{w}^{(s)}(x_{2}, 0)]+\cdots+\beta_{2N}[\tilde{w}^{(f)}(x_{N}, 0)+\tilde{w}^{(s)}(x_{N}, 0)] \\
     \quad \quad \vdots  \\
     \tilde{w}^{(s)}(x_{N}, 0) =\beta_{N1}[\tilde{w}^{(f)}(x_{1}, 0)+\tilde{w}^{(s)}(x_{1}, 0)]+\beta_{N2}[\tilde{w}^{(f)}(x_{2}, 0)+\tilde{w}^{(s)}(x_{2}, 0)]+\cdots+\beta_{NN}[\tilde{w}^{(f)}(x_{N}, 0)+\tilde{w}^{(s)}(x_{N}, 0)] \\
   \end{cases}
\end{equation}
\end{small}
\noindent
where:
\begin{equation} \label{equ:beta_mn}
\begin{split}
    \beta_{nj}&=\frac{-\Omega_j}{2\pi\mu} \frac{1}{2a} \int_{-\infty}^{\infty}\frac{k_s^{2} p}{R(k)} \frac{\sin(ka)}{ka}\,\mathrm{d}k \int_{x_n-a}^{x_n+a} \mathrm{e}^{\mathrm{i}k(x-x_j)}\,\mathrm{d}x \\
    &=\frac{-\Omega_j}{2\pi\mu} \int_{-\infty}^{\infty}\frac{k_s^{2} p}{R(k)} \frac{\sin^2(ka)}{(ka)^2} \mathrm{e}^{\mathrm{i}k(x_n-x_j)}\,\mathrm{d}k  \quad \text{for} \quad n,j=1,2,\cdots,N
\end{split}
\end{equation}

For simplicity, we express Eq. (\ref{equ:expanding vertical scattered field}) in matrix form as:

\begin{equation} \label{equ:matrix form}
    \bm{A} \bm{x}=\bm{b}
\end{equation}
\noindent
where the corresponding coefficients are:

\begin{subequations}
\begin{equation}
    \bm{A} = \left[\begin{array}{cccc}
    {(1-\beta_{11})} & {-\beta_{12}} & {\cdots} & {-\beta_{1N}} \\
    {-\beta_{21}} & {(1-\beta_{22})} & {\cdots} & {-\beta_{2N}} \\
    {\vdots} & {\vdots} & {\ddots} & {\vdots} \\
    {-\beta_{N1}} & {-\beta_{N2}} & {\cdots} & {(1-\beta_{NN})} \\
    \end{array}\right] \in \mathbb{C}^{N \times N}
\end{equation}

\begin{equation}
    \bm{x}=\left[\begin{array}{c}
    {\tilde{w}^{(s)}(x_{1}, 0)} \\
    {\tilde{w}^{(s)}(x_{2}, 0)} \\
    \vdots \\
    {\tilde{w}^{(s)}(x_{N}, 0)} \\  
    \end{array}\right] \in \mathbb{C}^{N \times 1}, \quad 
    \bm{b}=\left[\begin{array}{c}
    {\beta_{11}\tilde{w}^{(f)}(x_{1}, 0)+\beta_{12}\tilde{w}^{(f)}(x_{2}, 0)+\cdots+\beta_{1N}\tilde{w}^{(f)}(x_{N}, 0)} \\
    {\beta_{21}\tilde{w}^{(f)}(x_{1}, 0)+\beta_{22}\tilde{w}^{(f)}(x_{2}, 0)+\cdots+\beta_{2N}\tilde{w}^{(f)}(x_{N}, 0)} \\
    \vdots \\
    {\beta_{N1}\tilde{w}^{(f)}(x_{1}, 0)+\beta_{N2}\tilde{w}^{(f)}(x_{2}, 0)+\cdots+\beta_{NN}\tilde{w}^{(f)}(x_{N}, 0)} \\  
    \end{array}\right] \in \mathbb{C}^{N \times 1}
\end{equation}
\end{subequations}
The solution of Eq. (\ref{equ:matrix form}) provides the sought average vertical displacement components of the scattered field at the resonator footprint locations $x_n$. When the matrix $\bm{A}$ has a nonzero determinant, the system in Eq. (\ref{equ:matrix form}) has a unique solution $\bm{x}$ with components:

\begin{equation} \label{equ:solution of scattered field at x_j}
    \tilde{w}^{(s)}(x_n, 0)=(\bm{A}^{-1} \bm{b})_n \quad \text{for} \quad n=1,2,\cdots,N
\end{equation}
\noindent
At this stage, by substituting Eq. (\ref{equ:solution of scattered field at x_j}) into Eqs. (\ref{equ:scattered u by the j-th resonator}), (\ref{equ:scattered w by the j-th resonator}), we can obtain the $j$-th scattered field components $u_j^{(s)}(x,z)$ and $w_j^{(s)}(x,z)$ at any point of the $x-z$ plane. The total wavefield is then obtained as the summation of the free and scattered fields:
\begin{subequations}
\begin{equation} \label{equ:horizontal total field expression}
    u(x,z)=u^{(f)}(x,z)+u^{(s)}(x,z)=u^{(f)}(x,z)+\sum_{j=1}^{N} u_{j}^{(s)}(x,z)
\end{equation}
\begin{equation} \label{equ:vertical total field expression}
    w(x,z)=w^{(f)}(x,z)+w^{(s)}(x,z)=w^{(f)}(x,z)+\sum_{j=1}^{N} w_{j}^{(s)}(x,z)
\end{equation}
\end{subequations}
\noindent

Equations \eqref{equ:horizontal total field expression} and \eqref{equ:vertical total field expression} can be rewritten in integral form as follows:

\begin{small}
\begin{equation} \label{equ:horizontal total field}
\begin{split}
    u(x,z) &=\frac{\mathrm{i}Q}{2\pi\mu} \int_{-\infty}^{\infty}\frac{k(2k^2-k_s^{2}) \mathrm{e}^{pz}-2kpq \mathrm{e}^{qz}}{R(k)} \mathrm{e}^{\mathrm{i}kx}\,\mathrm{d}k \\
    & + \frac{\mathrm{i}}{2\pi\mu} \sum_{j=1}^{N} \Omega_{j}\left[\frac{-Q}{2\pi\mu} \int_{-\infty}^{\infty}\frac{k_s^{2} p}{R(k)} \frac{\sin(ka)}{ka} \mathrm{e}^{\mathrm{i}kx_j}\,\mathrm{d}k + (\bm{A}^{-1} \bm{b})_j\right]\int_{-\infty}^{\infty}\frac{k(2k^2-k_s^{2}) \mathrm{e}^{pz}-2kpq \mathrm{e}^{qz}}{R(k)} \frac{\sin(ka)}{ka} \mathrm{e}^{\mathrm{i}k(x-x_j)}\,\mathrm{d}k
\end{split}
\end{equation}
\end{small}

\begin{small}
\begin{equation} \label{equ:vertical total field}
\begin{split}
    w(x,z) &=\frac{Q}{2\pi\mu} \int_{-\infty}^{\infty}\frac{p(2k^2-k_s^{2}) \mathrm{e}^{pz}-2k^2p \mathrm{e}^{qz}}{R(k)} \mathrm{e}^{\mathrm{i}kx}\,\mathrm{d}k \\
    & +\frac{1}{2\pi\mu} \sum_{j=1}^{N} \Omega_{j}\left[\frac{-Q}{2\pi\mu} \int_{-\infty}^{\infty}\frac{k_s^{2} p}{R(k)} \frac{\sin(ka)}{ka} \mathrm{e}^{\mathrm{i}kx_j}\,\mathrm{d}k + (\bm{A}^{-1} \bm{b})_j\right]\int_{-\infty}^{\infty}\frac{p(2k^2-k_s^{2}) \mathrm{e}^{pz}-2k^2p \mathrm{e}^{qz}}{R(k)} \frac{\sin(ka)}{ka} \mathrm{e}^{\mathrm{i}k(x-x_j)}\,\mathrm{d}k
\end{split}
\end{equation}
\end{small}

The above Eqs. \eqref{equ:horizontal total field} and \eqref{equ:vertical total field} can be specified at the coordinate $z=0$ to obtain the surface displacement components:

\begin{equation} \label{equ:horizontal total field at z=0}
\begin{split}
    u(x, 0) &=\frac{\mathrm{i}Q}{2\pi\mu} \int_{-\infty}^{\infty}\frac{k(2k^2-k_s^{2}-2pq)}{R(k)} \mathrm{e}^{\mathrm{i}kx}\,\mathrm{d}k \\
    & + \frac{\mathrm{i}}{2\pi\mu} \sum_{j=1}^{N} \Omega_{j}\left[\frac{-Q}{2\pi\mu} \int_{-\infty}^{\infty}\frac{k_s^{2} p}{R(k)} \frac{\sin(ka)}{ka} \mathrm{e}^{\mathrm{i}kx_j}\,\mathrm{d}k + (\bm{A}^{-1} \bm{b})_j\right]\int_{-\infty}^{\infty}\frac{k(2k^2-k_s^{2}-2pq)}{R(k)} \frac{\sin(ka)}{ka} \mathrm{e}^{\mathrm{i}k(x-x_j)}\,\mathrm{d}k
\end{split}
\end{equation}

\begin{equation} \label{equ:vertical total field at z=0}
\begin{split}
    w(x, 0) &=\frac{-Q}{2\pi\mu} \int_{-\infty}^{\infty}\frac{k_s^{2} p}{R(k)} \mathrm{e}^{\mathrm{i}kx}\,\mathrm{d}k \\
    & -\frac{1}{2\pi\mu} \sum_{j=1}^{N} \Omega_{j}\left[\frac{-Q}{2\pi\mu} \int_{-\infty}^{\infty}\frac{k_s^{2} p}{R(k)} \frac{\sin(ka)}{ka} \mathrm{e}^{\mathrm{i}kx_j}\,\mathrm{d}k + (\bm{A}^{-1} \bm{b})_j\right]\int_{-\infty}^{\infty}\frac{k_s^{2} p}{R(k)} \frac{\sin(ka)}{ka} \mathrm{e}^{\mathrm{i}k(x-x_j)}\,\mathrm{d}k
\end{split}
\end{equation}

The closed-form Eqs. \eqref{equ:horizontal total field},  \eqref{equ:vertical total field},  \eqref{equ:horizontal total field at z=0},  \eqref{equ:vertical total field at z=0} are evaluated numerically via Gauss-Kronrod quadrature. In the next section we use the developed approach to predict the wavefield of half-spaces with different sets of resonators including uniform, graded and disordered arrays.

\section{Case studies} \label{Case studies}

In this section, we discuss the behavior of a single resonator, a couple of resonators, and a graded array of resonators placed atop an elastic half-space and excited by a far field source. Our aim is twofold: (i) to test and validate the accuracy of our formulation against a numerical (Finite Element) solution; (ii) to discuss the coupling between free and scattered wavefields for different mechanical parameters and layout of the resonators. \par

Before tackling these tasks, let us provide some useful parameters to facilitate our calculations and ease the further  discussions:

\begin{itemize}

  \item to generalize our conclusions, we introduce the  dimensionless mass parameter $\hat{m}_j = m_j/(\rho S \lambda_{rj})$, with $\rho$ being the density of the half-space and $\lambda_{rj}=c_R/f_{rj}$ the Rayleigh wavelength at resonant frequency $f_{rj}$; this parameter compares the mass of the resonator  to the mass of a portion of substrate under the resonator footprint area, with volume  $V=S\times \lambda_{rj}$. Additionally, we introduce the parameter $\hat{d}_{mn}=d_{mn}/\lambda_{rj}=(x_n-x_m)/\lambda_{rj}$ to denote the normalized distance between the $m$-th and $n$-th resonator.

  \item to account for wave energy dissipation, we assume a non-null damping ratio $\zeta$ for the resonator response and a non-null hysteretic damping $\xi$ for the elastic substrate. The damping coefficient of the $j$-th resonator is calculated as $c_j=2 m_j \omega_{rj} \zeta$ \cite{chopra2012dynamics}, while the complex moduli of the substrate as $\lambda'=\lambda(1+2\mathrm{i}\xi)$ and $\mu'=\mu(1+2\mathrm{i}\xi)$ \cite{philippacopoulos1988lamb}; the introduction of damping in the system allows also to avoid numerical instabilities by removing the poles of the integrands in Eqs. (\ref{equ:horizontal total field at z=0}, \ref{equ:vertical total field at z=0});

    \item to quantify the  contribution of the resonators scattered field at the half-space surface ($z=0$), we introduce the following amplitude ratio ($A_R$) \cite{woods1968screening}:
    \begin{equation} \label{equ:definition of transmission}
    A_R = \frac{w(x, 0)}{w^{(f)}(x, 0)} = 1+\frac{w^{(s)}(x, 0)}{w^{(f)}(x, 0)} 
    \end{equation}
 and the normalized distance between the receiver and the first resonator $\hat{d}=d/\lambda_{r1}=(x-x_1)/\lambda_{r1}$ where $A_R$ will be computed.
    \end{itemize}
    
\subsection{Single resonator scenario} \label{Single resonator scenario}

We begin our investigation considering the case of a single surface oscillator. 
The resonator has a normalized mass $\hat{m}_1=1$, is located at a distance $x_1=6\lambda_{r1}$ from the harmonic source, and lies over a half-space characterized by the mechanical properties  collected in Table \ref{tab:tab1}.

\begin{figure}[hbt!]
    \centering
    \includegraphics[width=6.5 in]{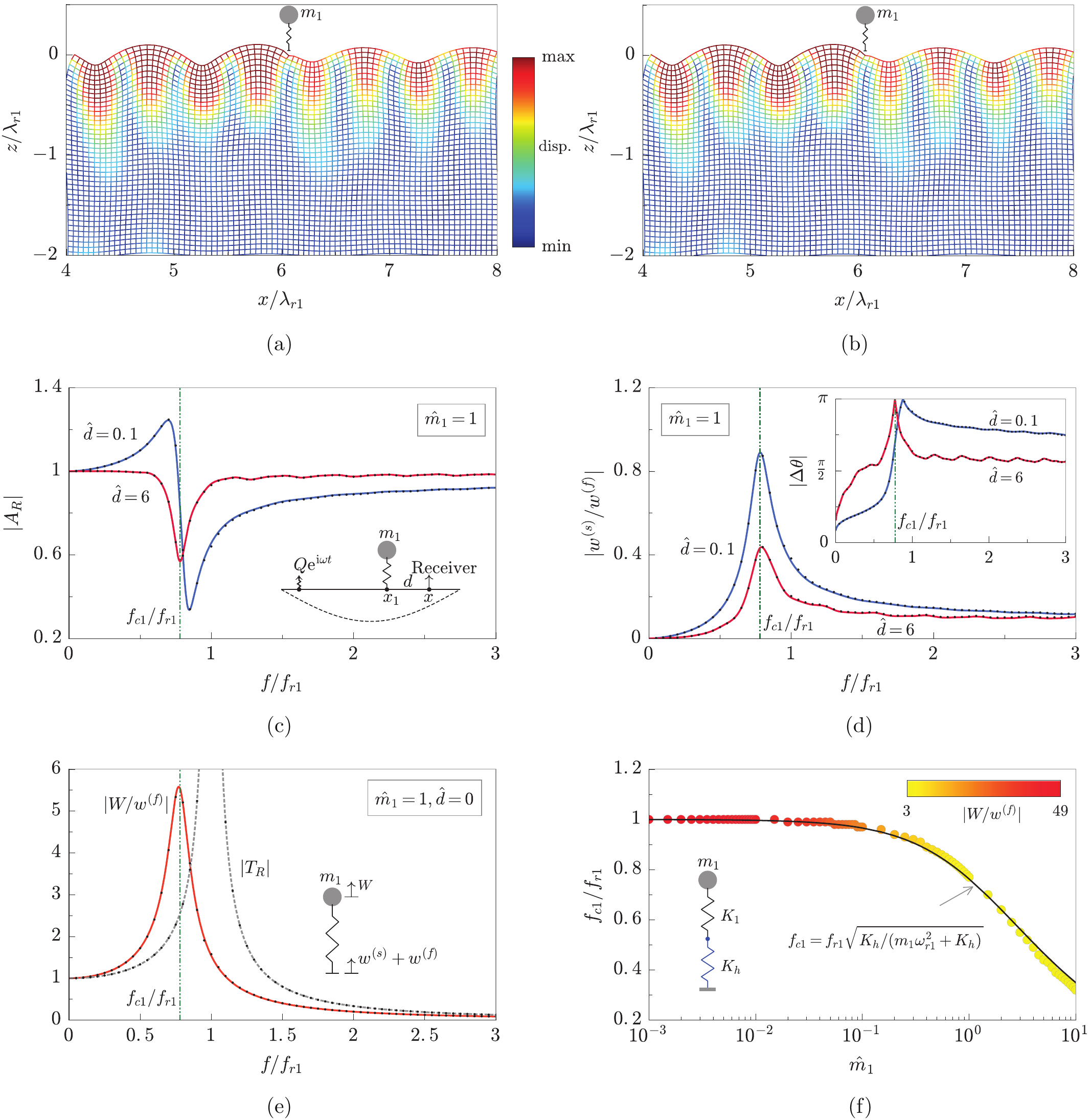}
    \caption{Rayleigh wave interaction with a single resonator on an elastic half-space ($\hat{m}_1=1$, $x_1 = 6 \lambda_{r1}$). Total wavefield for a harmonic source at $f=f_{r1}$ computed using the (a) proposed analytical solution, (b) FE simulation. (c) Amplitude ratio $|A_R|$ in the near $\hat{d}=0.1$ (blue line) and far field $\hat{d}=6$ (red line). (d) Spectrum and phase of the scattered field vs. the free field. (e) Dynamic amplification factor of the resonator on a rigid (dashed line) and elastic substrate (red line). (f) Coupled frequency $f_{c1}$ vs. the mass of the resonator. For comparison, in (c), (d) and (e) we also provide the FE solutions denoted by dots.}
    \label{fig:fig2}
\end{figure}

The total wavefield generated by a harmonic source at $f=f_{r1}$, computed by using Eqs. \eqref{equ:horizontal total field} and \eqref{equ:vertical total field}, is shown in Fig. \ref{fig:fig2}a in the domain $x=[4, 8]\lambda_{r1}$, $z=[-2, 0]\lambda_{r1}$. 
The analytically predicted wavefield is in excellent agreement with the one displayed in Fig. \ref{fig:fig2}b, calculated with harmonic FE simulations (see \ref{Appendix} for details on the FE model). 
To quantify the effect of the resonator on the surface wavefield, we calculate and show in Fig. \ref{fig:fig2}c the amplitude ratio $|A_R|$ for harmonic sources with frequencies $f=[0, 3]f_{r1}$ at two positions, namely $\hat{d}=0.1$ (blue line), in the scattered near field, and at $\hat{d}=6$ (red line), in the scattered far field.

\begin{table}[H]
\caption{Mechanical parameters of the resonators and the elastic half-space. The elastic parameters are taken from Ref. \cite{pu2020broadband}.}
\label{tab:tab1}
\begin{tabular*}{\hsize}{@{}@{\extracolsep{\fill}}ll@{}}
\toprule
Parameter & Value  \\   
\midrule
Line load amplitude, $Q$                         & $10^6$ N/m \\
First resonator frequency, $f_{r1}$              & 2 Hz \\
Footprint length of each resonator, $2a$             & 1 m  \\
Damping ratio of resonators, $\zeta$             & 1$\%$ \\
Mass density of substrate, $\rho$                & 1800 kg/m$^3$ \\
Young modulus of substrate, $E$                  & 46 MPa \\
Poisson ratio of substrate, $\nu$                & 0.25 \\
Hysteretic damping ratio of substrate, $\xi$     & 1$\%$ \\
\bottomrule
\end{tabular*}
\end{table}

In the near field response we observe a low-frequency region where the surface motion is amplified with respect to the free field, namely $|A_R|>1$, followed by a higher frequency region with a significant deamplification of the signal. The two regimes are separated by the frequency $f_{c1}$ which corresponds to the resonance of the oscillator coupled to the elastic substrate. In the far field the signal amplification disappears and we observe only a deamplification of the signal with a maximum drop occurring exactly at the coupled frequency $f_{c1}$. \par  

To interpret the different responses observed near and far from the resonator, we expand the expression of $|A_R|$ given in Eq. (\ref{equ:definition of transmission}) as:

\begin{equation} \label{equ:transmission of one resonator}
    |A_R| = \sqrt{1+\left|\frac{w^{(s)}}{w^{(f)}}\right|^2+2 \mathrm{Re} \left(\frac{w^{(s)}}{w^{(f)}}\right)} = \sqrt{1+\left|\frac{w^{(s)}}{w^{(f)}}\right|^2+2\left|\frac{w^{(s)}}{w^{(f)}}\right|\cos(\Delta \theta)}
\end{equation}
\noindent
where $\Delta \theta$ is the relative phase between the scattered and free  vertical displacements, namely, $w^{(s)}/w^{(f)}$. \par

According to the Eq. (\ref{equ:transmission of one resonator}), both the amplitude and the relative phase of the scattered field play a role in the amplitude ratio $|A_R|$: the scattered field amplitude shows a similar trend both in the near and in the far field and reaches its peak exactly at the coupled frequency $f_{c1}$ (see Fig. \ref{fig:fig2}d); conversely, the relative phase $\Delta\theta$ significantly changes depending on the receiver location. In the near field, around the coupling frequency, the relative phase (see inset in Fig. \ref{fig:fig2}d) shows that the scattered and free responses are approximately orthogonal. Hence, the contribution of the scattered waves to the value of $|A_R|$ is negligible. Conversely, in the far field, scattered and free responses are out-of-phase, leading to destructive interference between the two wavefields and to a minimum value of the amplification ratio $|A_R|$. \par

Let us now investigate in more detail the variation of the coupled frequency $f_{c1}$ with respect to the mechanical parameters of the system, namely substrate and oscillator. To this purpose, we calculate the absolute vertical displacement $W$ of a resonator located on the half-space surface at distance $d=0$ from the source and compare its response to the one of an identical resonator placed on a rigid substrate. The two resonators vertical responses, normalized with respect to the relative base displacements, are plotted in Fig. \ref{fig:fig2}e. As expected, the response of the resonator on a rigid substrate provides the transmissibility factor $|T_R|$ (dashed line), whereas the maximum amplitude response of the resonator lying on the half-space (red continuous line) does not occur at its natural frequency $f_{r1}$ but is shifted towards the coupled resonant frequency $f_{c1}$. This frequency shift can be predicted utilizing a lumped-mass model, in which the contribution of the half-space is lumped in the additional stiffness $K_h$ (see schematic in Fig. \ref{fig:fig2}f). Hence, the coupled frequency can be calculated as:
\begin{equation} \label{equ:equivalent coupled frequency}
     f_{c1}= f_{r1}\sqrt{\frac{K_h}{m_1 \omega_{r1}^2+K_h}}
\end{equation}
\noindent
where the equivalent substrate stiffness $K_h$ can be estimated from Eq. (\ref{equ:scattered w by the j-th resonator}) by considering the averaged stress over the contact area ($z=0$) and a null source-response distance ($x=x_j$):
\begin{equation} \label{equ:equivalent substrate stiffness}
     \frac{1}{K_{h}}= \frac{-1}{2\pi\mu} \int_{-\infty}^{\infty}\frac{k_s^{2} p}{R(k)} \frac{\sin^2(ka)}{(ka)^2}\,\mathrm{d}k
\end{equation}
\noindent
Note that the integral in Eq. (\ref{equ:equivalent substrate stiffness}) is frequency dependent, so strictly speaking the coupling frequency $f_{c1}$ in Eq. (\ref{equ:equivalent coupled frequency}) is frequency dependent too. As an approximation, here we calculate the value of $K_h$ at $f=f_{r1}$ and use it to predict the value of $f_{c1}$ for different resonator mass. \par

The value of the coupled frequency vs. the resonator mass as predicted from Eq. (\ref{equ:equivalent coupled frequency}) is plotted in Fig. \ref{fig:fig2}f as a continuous black line. The prediction agrees well with the frequency values at which the response of the different resonators reach its maximum amplitude as calculated from Eq. \eqref{equ:relation of displacement} (marked in Fig. \ref{fig:fig2}f by colored circles). In particular, the color of the circles shows the value of the amplification factor ($W/w^{(f)}$) at resonance, and highlights that a change of mass produces a change in the effective quality factor of the resonator too. To conclude this section, we remark that all these results have been validated with FE models (see black dots superimposed onto all the curves in Figs. \ref{fig:fig2}c,d,e).

\subsection{Two resonators scenario} \label{Two resonators scenario}

Studies on the dynamics of coupled oscillators on elastic supports are receiving renovate attention both in the geophysical context, to assess the response of close buildings and their influence on ground vibrations, and in the design of SAW devices, where micro/nano resonators are proposed for application in classical and quantum information processing \cite{raguin2019dipole}. In what follows, we show how our analytical formulation allows to properly analyze the mutual interaction between close resonators and to quantify its effects on the half-space and resonator responses.\par 

\begin{figure}[hbt!]
    \centering
    \includegraphics[width=6.5 in]{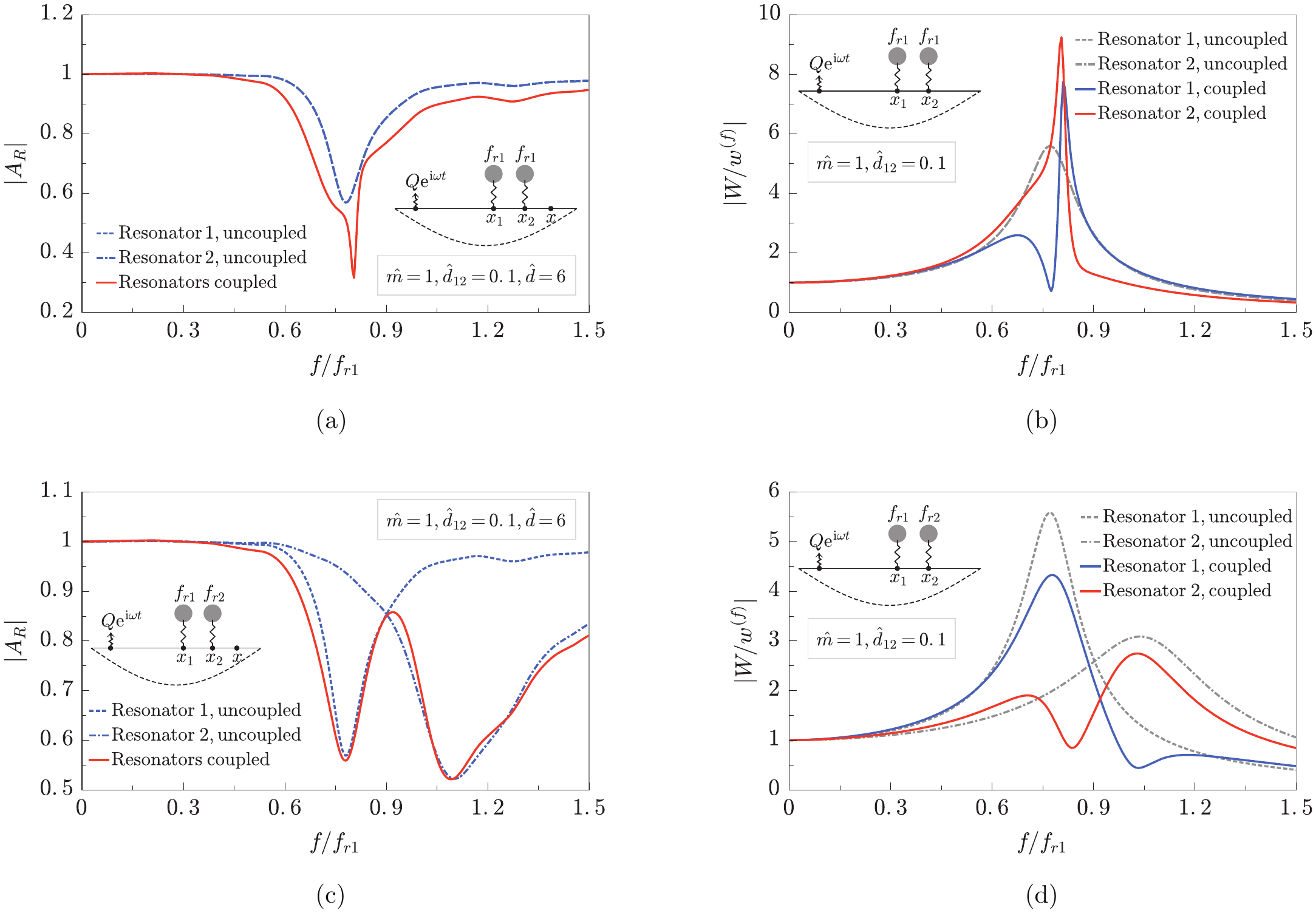}
    \caption{Rayleigh wave interaction with two resonators on an elastic half-space ($\hat{m}_1=\hat{m}_2=1, x_1=6\lambda_{r1}, d_{12}=0.1\lambda_{r1}$). (a) Amplitude ratio $|A_R|$ computed at $\hat{d}=6$ for two identical resonators ($f_{r2}=f_{r1}$). (b) Uncoupled and coupled amplification factors for two identical resonators ($f_{r2}=f_{r1}$). (c) Amplitude ratio $|A_R|$ computed at $\hat{d}=6$ for two different resonators ($f_{r2}=2f_{r1}$). (d) Uncoupled and coupled amplification factors for two different resonators ($f_{r2}=2f_{r1}$).}
    \label{fig:fig3}
\end{figure}

We begin our investigation considering two identical resonators, with parameters given in Table \ref{tab:tab1} and with a relative spacing $d_{12}=0.1\lambda_{r1}$, excited by a far field ($x_1=6\lambda_{r1}$) harmonic source. We calculate the amplitude ratio $|A_R|$ for a receiver located at $\hat{d}=6$ from the first resonator. The presence of two oscillators atop the half-space leads to a significant reduction in the amplitude ratio which shows its minimum value at a frequency different from the resonator coupled frequency ($f_{c1}=f_{c2}$). We remark that the system response cannot be predicted from the simple superposition of the single resonator scenarios,  which would neglect the cross-coupling interaction between the two resonators (see the $|A_R|$ of each single resonator scenario (blue lines) in Fig. \ref{fig:fig3}a).
To investigate further the coupling between the responses of the resonators, we expand the expression of the amplitude ratio as: 
\begin{equation} \label{equ:transmission of two resonators}
    |A_R| = \sqrt{1 + \left|\frac{w_1^{(s)}}{w^{(f)}}\right|^2 + \left|\frac{w_2^{(s)}}{w^{(f)}}\right|^2 + 2\left|\frac{w_1^{(s)}}{w^{(f)}}\right|\cos(\Delta \theta_1) + 2\left|\frac{w_2^{(s)}}{w^{(f)}}\right|\cos(\Delta \theta_2) + 2\left|\frac{w_1^{(s)}}{w^{(f)}}\right| \left|\frac{w_2^{(s)}}{w^{(f)}}\right| \cos(\Delta \theta_2-\Delta \theta_1)}
\end{equation}
\noindent
where $\Delta \theta_1$ and $\Delta \theta_2$ denote the phase of $w_1^{(s)}/w^{(f)}$ and $w_2^{(s)}/w^{(f)}$, respectively. The above equation clearly highlights the presence of the cross-coupling contribution, namely, the last term in Eq. (\ref{equ:transmission of two resonators}), in addition to the independent contributions of each resonator. \par

This mutual interaction modifies also the resonators responses, as highlighted by Fig. \ref{fig:fig3}b where the uncoupled and coupled oscillators amplification factors are shown. In particular, by looking at the maximum of the $|A_R|$ factor, one can observe a shift in the resonance of both oscillators. Additionally, the first resonator shows a characteristic ``frequency splitting" behavior, recently observed experimentally in couples of micropillars attached to an elastic substrate \cite{raguin2019dipole}. \par

An effective mutual interaction requires resonators with similar (possibly identical) resonance frequencies located at a relatively short distance. That is because the scattered fields are maximized at the coupled resonance frequencies and in the near fields of the resonators, as shown in Fig. \ref{fig:fig2}d for a single resonator. To evidence this effect, let us consider a configuration of two resonators with different natural frequencies, e.g., $f_{r2}=2f_{r1}$, keeping the other parameters unchanged. Fig. \ref{fig:fig3}c shows the amplitude ratio at $\hat{d}=6$. The reader can appreciate that the system response is now simply the envelope, namely the superposition of the two single case scenarios. Similarly, the amplification factors of the two resonators, reported in Fig. \ref{fig:fig3}d, confirm that no significant shift occurs in the coupled frequencies of each resonator. \par

We remark that the extraction of these features is eased by our analytical framework which allows to investigate and distinguish the contribution of each scattered field to the total response. 

\subsection{Cluster of resonators} \label{Cluster of resonators}

So far we have gained some physical insights into the behavior of single and coupled oscillators lying over an elastic substrate. When the number of resonators increases to form a cluster, or a so-called metasurface, the propagation of surface waves in the elastic substrate can be characterized by intriguing phenomena,  such as surface-to-bulk wave conversions (from classical \cite{boechler2013interaction} and Umklapp scattering \cite{chaplain2020tailored}) and wave localization  (via classical \cite{colombi2016seismic} and topological rainbow trapping \cite{ chaplain2020topological}).  Although the extraction of the dispersive properties of such graded clusters can be typically inferred from analytical models developed for infinite regular  arrays \cite{colquitt2017seismic}, the evaluation of the related wavefields require the use of numerical schemes. In what follows, we will show that our analytical framework can properly capture these phenomena. \par

\begin{figure}[hbt!]
    \centering
    \includegraphics[width=6.5 in]{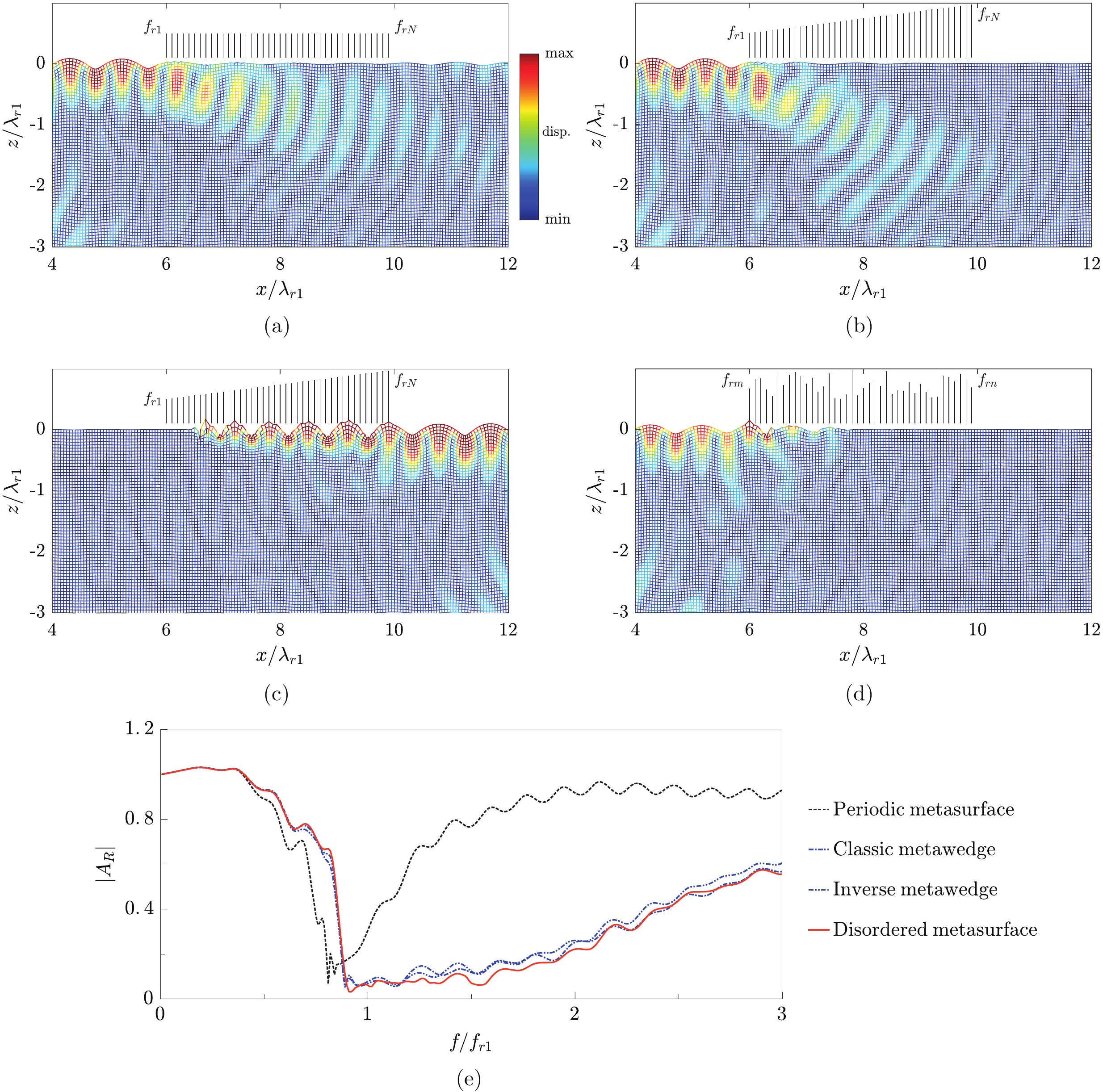}
    \caption{Rayleigh wave interaction with finite-size metasurfaces ($N=40$) on an elastic half-space ($\hat{m}=0.5$, resonator spacing $0.1\lambda_{r1}$, incident frequency $f=1.1f_{r1}$). Calculated  wave field for: (a) a periodic metasurface. (b) an inverse metawedge. (c) a classical metawedge. (d) a disordered metasurface. In the schematics the length of the resonators represent the value of the related natural frequencies. (e) Amplitude ratio $|A_R|$ computed at $\hat{d}=6$ in the frequency range $f=[0, 3]f_{r1}$ of the four considered cases.}
    \label{fig:fig4}
\end{figure}

We begin by considering an array of 40 identical resonators ($\hat{m}_j=0.5, f_{rj}=f_{r1}$), arranged periodically with a lattice spacing  $0.1\lambda_{r1}$ (see Fig. \ref{fig:fig4}a). Similar arrays have been analyzed to assess the surface wave filtering capabilities of a periodic metasurface \cite{colombi2016seismic,pu2020seismic}. For such a configuration, we expect the existence of a band gap in a narrow frequency region above the oscillator resonance and the occurrence of surface-to-shear wave conversion. \par

To visualize this phenomenon, we consider an incident Rayleigh wave with frequency $f=1.1f_{r1}$, namely in the metasurface band gap, which is excited by the harmonic source $Q\mathrm{e}^{\mathrm{i}\omega t}$ sufficiently far away from the metasurface ($x_1=6\lambda_{r1}$). The displacement wavefield is displayed in Fig. \ref{fig:fig4}a and shows how the incident Rayleigh wave is converted into a downward propagating shear wave. \par

The same wave-conversion phenomenon can be extended over a broader frequency range by utilizing a series of resonators with a graded variation of frequency along the array \cite{colombi2016seismic}. To achieve this purpose, we model an array of 40 resonators ($\hat{m}_j=0.5$) with resonant frequencies linearly increasing along the array with a step $\Delta f_{r}=0.05f_{r1}$, for an overall working bandwidth between $f_{r1}$ and 3$f_{r1}$. Alike the periodic metasurface previously discussed, we set a constant spacing between the resonators. Indeed, this graded configuration, also known in the literature as metawedge \cite{colombi2016seismic}, supports both a mode conversion and a wave localization, depending on the direction of the incoming excitation. These two distinct wavefields, as calculated using our analytical framework, are shown respectively in Figs. \ref{fig:fig4}b and \ref{fig:fig4}c. \par

As last example, we calculate the wavefield of a disordered metasurface, constructed by randomly changing the order of resonators considered in the discussed metawedge. The investigation of such a non-periodic resonant system is attracting increasing interest in the research community as a design strategy to enlarge the filtering bandwidth of metamaterials \cite{celli2019bandgap, cao2020disordered}. Note that a random configuration can be easily modeled using the proposed analytical approach which allows to set frequency and location of the resonating scatters at will. In Fig. \ref{fig:fig4}d the reader can appreciate the related wavefield for an incident Rayleigh wave at $f=1.1f_{r1}$. \par

We also provide a comparison of the amplitude ratio $|A_R|$ calculated at $\hat{d}=6$ in the frequency range $f=[0, 3]f_{r1}$. The result is shown in Fig. \ref{fig:fig4}e and highlights that the disordered metasurface provides  similar filtering performance to the two analyzed graded systems. 

\section{Conclusions} \label{Conclusions}

This work proposes an analytical formulation to model and study the interaction of vertically polarized elastic waves with surface resonators. In particular, we have exploited the Green's functions of the canonical Lamb's problem to setup a coupled problem between the incident field generated by the source and the scattered wavefields generated by a set of resonators placed atop the half-space. The formulation can handle an arbitrary number of resonators arranged on the surface of the half-space in a generic configuration. The capabilities of the developed methodology have been discussed by modeling the dynamics of a single, a couple and a cluster of resonators arranged over an isotropic homogeneous half-space, and have been validated against finite element simulations. \par

The method allows to capture the frequency shift of a resonator coupled to the elastic substrate, the mutual interaction between a couple of close resonators in terms of frequency splitting and amplitude variation and the collective response of arrays of resonators, e.g., metasurfaces interacting with Rayleigh waves. Future research efforts will be devoted to extend the methodology to a 3D scenario and exploit its capability to design SAW devices, waveguides and interpret the dynamics of a cluster of buildings interacting with seismic waves and urban vibrations.

\section*{CRediT authorship contribution statement}
\noindent \textbf{Xingbo Pu:} Conceptualization, Methodology, Investigation, Software, Data curation, Writing - original draft. \textbf{Antonio Palermo:} Conceptualization, Investigation, Validation, Writing - review \& editing, Co-supervision. \textbf{Alessandro Marzani:} Conceptualization, Investigation, Writing - review \& editing, Supervision, Funding acquisition.

\section*{Declaration of competing interest}
\noindent The authors declare that they have no conflict of interest.

\section*{Acknowledgments}
\noindent This project has received funding from the European Union’s Horizon 2020 research and innovation programme under the Marie Skłodowska Curie grant agreement No 813424. A.P. acknowledges the support of the University of Bologna - DICAM through the research fellowship “Metamaterials for seismic waves attenuation”. 

\appendix
\section{Details on the FE model} \label{Appendix}
In this Appendix, we provide details of the FE model (see Fig. \ref{fig:figA1}) used to validate our analytical solutions in Section \ref{Single resonator scenario}. First, to simulate the uniform vertical force imposed on the footprint, the mass-dashpot-spring oscillator is discretized as an ensemble of 11 truss elements, namely the truss spacing $0.2a \ll \lambda_r$. This procedure results in each point mass $m_p = m/11$, and the Young modulus of each truss $E_t = (m_p \omega_r^2+\mathrm{i} \omega c)/A_t$, where $A_t$ is the cross-sectional area of a truss. The incident Rayleigh wave is excited by the harmonic load with amplitude $Q$ at a sufficient distance ($6\lambda_r$) from the oscillator. To model the elastic half-space and to avoid unnecessary reflections, we add Perfectly Matched Layers (PMLs) to the vertical and bottom edges.

\setcounter{figure}{0}
\begin{figure}[hbt!]
    \centering
    \includegraphics[width=5 in]{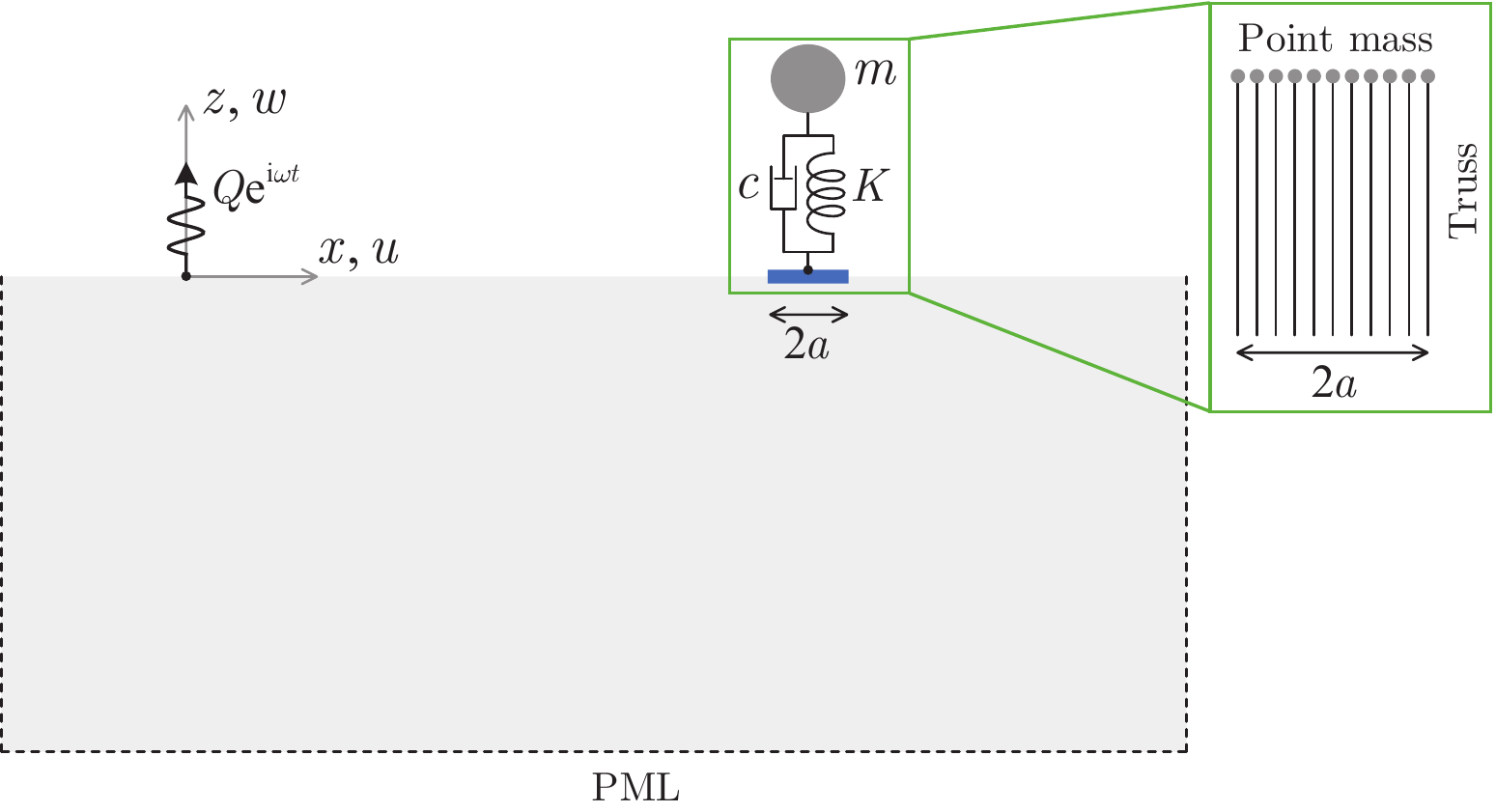}
    \caption{Schematic of FE model.}
    \label{fig:figA1}
\end{figure}

\bibliography{mybibfile}

\end{document}